\documentstyle[12pt,aaspp4]{article}
\newcommand{\dmm}{\mbox{$\Delta$m$_{15}(B)$}}
\newcommand{\kms}{\mbox{km s$^{-1}$}}
\newcommand{\ho}{\mbox{$H_\circ$}}
\begin{document}

\title{Optical Light Curve of the Type Ia Supernova 1998bu in M96 and
the Supernova Calibration of the Hubble Constant}

\author{
Nicholas B. Suntzeff
}
\affil{
Cerro Tololo Inter-American
Observatory, National Optical Astronomy
Observatories,\altaffilmark{1}\\ Casilla 603, La Serena, Chile 
}
\and
\author{M. M. Phillips
}
\affil{Las Campanas Observatory, Carnegie Observatories, Casilla 601,
La Serena, Chile
}
\and
\author{
R. Covarrubias, M. Navarrete, J. J. P\'erez, A. Guerra, M. T. Acevedo
}
\affil{
Cerro Tololo Inter-American
Observatory, National Optical Astronomy
Observatories,\altaffilmark{1}\\ Casilla 603, La Serena, Chile 
}
\and
\author{Laurance R. Doyle
}
\affil{SETI Institute, MS 245-3, NASA Ames Research Center, 
Moffett Field, CA  94035
}
\and
\author{Thomas Harrison
}
\affil{New Mexico State Univ., Box 30001/Dept. 4500, Las Cruces, NM 88003
}
\and
\author{Stephen Kane, Knox S. Long
}
\affil{STScI, 3700 San Martin, Dr., Baltimore, MD 21218
}
\and
\author{Jos\'e Maza}
\affil{
Dept. Astronom\'ia,
Universidad de Chile,
Casilla 36-D,
Santiago, Chile\\
}
\and
\author{Scott Miller
}
\affil{Dept. of Astronomy, Univ. of Maryland, College Park, MD 20742
}
\and
\author{Andr\'es E. Piatti, Juan J. Clari\'a, Andrea V. Ahumada
}
\affil{Obs. Astron. C\'ordoba, Laprida 854, 5000, Cordoba, Argentina
}
\and
\author{Barton Pritzl
}
\affil{Dept. of Physics and Astronomy, Michigan State Univ., East
Lansing, MI 48824
}
\and
\author{P. Frank Winkler
}
\affil{Dept. of Physics, Middlebury College, Middlebury, VT 05753
}
\altaffiltext{1}{
Cerro Tololo Inter-American Observatory, National
Optical Astronomy Observatories, operated by the Association of Universities
for Research in Astronomy, Inc., (AURA), under cooperative agreement with
the National Science Foundation.
}
\authoremail{nsuntzeff@noao.edu}


\begin{abstract}

We present the $UBV(RI)_{KC}$ light curves of the Type~Ia supernova
SN~1998bu which appeared in the nearby galaxy M96 (NGC 3368). M96 is a
spiral galaxy in the Leo I group which has a Cepheid-based
distance. Our photometry allows us to calculate the absolute magnitude
and reddening of this supernova.  These data, when combined with
measurements of the four other well-observed supernovae with Cepheid
based distances, allow us to calculate the Hubble constant with
respect to the Hubble flow defined by the distant Cal\'an/Tololo Type
Ia sample. We find a Hubble constant of 64.0 $\pm$ 2.2(internal) $\pm$
3.5(external) \kms\ Mpc$^{-1}$, consistent with most previous
estimates based on Type~Ia supernovae. We note that the two
well-observed Type Ia supernovae in Fornax, if placed at the Cepheid
distance to the possible Fornax spiral NGC 1365, are apparently too
faint with respect to the Cal\'an/Tololo sample calibrated with the
five Type Ia supernovae with Cepheid distances to the host galaxies.

\end{abstract}

\section{Introduction}

When corrected for reddening and small luminosity differences, the
Type Ia supernova peak magnitudes show dispersion from a uniform
Hubble flow of only 0\fm12 for nearby supernovae with redshifts of $z
\sim 0.05$ (\cite{Rie_etal96,Phi_etal98}). The distance to a single Type
Ia supernova can therefore be measured to roughly 5\%, which makes
these objects very accurate distance probes for measuring the Hubble
constant (\cite{Ham_etal96b,Rie_etal96}) and the geometry of the
Universe (\cite{Per_etal97,Schm_etal98}). Measuring the geometry of
the Universe, which can be expressed in dimensionless constants,
requires the measurement of the relative magnitudes (corrected for
relativistic effects and K-corrections) for nearby and distant objects
(\cite{San61}). To measure the Hubble constant (which is not
dimensionless) with supernovae, we require a calibration of the
intrinsic luminosities of the objects.

If we ignore the ``second-order effects'' of reddening and the small
intrinsic luminosity differences between Type Ia supernovae, the
measurement of the Hubble constant requires three sets of observed
quantities to be accurately determined: (1) the true distance to a set
of Cepheids defining a period-luminosity relationship, often taken as
the distance to the LMC Cepheids, (2) the relative distance moduli
from the LMC to the galaxies hosting both Cepheids and a Type Ia
supernovae, and (3) the zero point of the observed Hubble diagram of
Type Ia supernovae that are distant enough to be in the undisturbed
Hubble flow.  The error in (1) is about 0\fm12 (\cite{Sil_etal98}):
however the range in values for the distance moduli to the LMC is much
larger at 18.3-18.7 (see \cite{Wal98} for a recent review). The error
in (3) is only 0\fm03 (\cite{Ham_etal96b}). The error in (2) for the
galaxy NGC 1365 studied by \cite{Sil_etal98} is 0\fm20. This galaxy is
more distant than the other calibrators and perhaps a better error
estimate is the average error for the 5 galaxies cited in this paper
(see Table \ref{t5}) or 0\fm15. This error should be combined with the
intrinsic dispersion in the peak magnitude for a single Type Ia
supernova, which is about 0\fm12 as mentioned above. Thus the final
error in (2) is $\sim 0\fm2/(n)^{0.5}$ where $n$ is the number of
galaxies in the sample.

With only four Type Ia supernovae with both Cepheid distances and
reliable light curves (\cite{Ham_etal96b}), the dominant source of
error in the Hubble constant based on supernovae is not from the the
zero point of the Hubble diagram of the distant sample, but from the
small number of supernovae in host galaxies with Cepheid distances. It
is therefore important to follow any supernova that appears in a
galaxy with a Cepheid distance. SN~1998bu which appeared in NGC 3368
(M96) was such a candidate. This galaxy in the Leo I group has a
Cepheid distance published by \cite{Tan_etal95}.

SN~1998bu was discovered on 9.9 May 1998 (\cite{Vil98}) (all dates
given as UT).  The coordinates of the supernova are (10:46:46.01,
+11:50:07.5, 2000.0) (\cite{Bos98}).  Both \cite{Aya_etal98} and
\cite{Mei_etal98} classified the supernova as a Type Ia before or near
maximum light based on spectra taken around 13 May.

\cite{Mun_etal98} obtained echelle spectra on 12 May which showed a
strong \ion{Na}{1} D interstellar absorption feature of 350m\AA\ (D1)
at $744.8 \pm 0.3$ \kms\ at the redshift of the host galaxy,
indicating a substantial reddening for the supernova in the host
galaxy.  They also found an interstellar component from the Galaxy at
$-6 \pm 1.5$ \kms\ with a \ion{Na}{1} (D1) equivalent width of
190m\AA. Using the \cite{MunZwi97} calibration, they estimated the
reddening due to the host galaxy and our Galaxy at 0.15 and 0.06 in
units of E($B-V$). The large interstellar absorption features were
confirmed by \cite{Cen_etal98} who found two components at the
\ion{Ca}{2} K with velocities of 743 and 750 \kms\ with equivalent
widths of 107 and 60 m\AA. They also found a weak \ion{Na}{1}
component at 722 \kms. The Galactic component of the \ion{Ca}{2} K
line was seen at five velocities: -26, -8, +7, +48, +53 \kms\ with
equivalent widths of 77, 154, 16, 12, 22 m\AA.

\cite{Sch98} used the photometry reported in the IAU Circulars and two
photometric data points from the WIYN telescope to estimate the time
of $B$ maximum of $21.0 \pm 0.3$ May with $\dmm = 0.95 \pm 0.05$ where
\dmm\ is the decline in $B$ magnitude over 15 days from maximum light
(\cite{Phi93}). He also estimated the peak magnitudes of $B = 12.36
\pm 0.05$ and $V = 11.93 \pm 0.04$.  \cite{SkiFar98} reported a
pre-discovery detection of $16.5 \pm 0.3$ on 3.14 May on an unfiltered
CCD image.

In the remaining sections of the paper, we discuss the observations
and reduction to the $UBVRI$ photometric system in \S\ref{s1}, and in
\S\ref{s2} we discuss the evaluation of the Hubble constant with the
addition of this new distance to a Type Ia supernova.

\section{Observations\label{s1}}

\subsection{Detector and Filters}

We began observations of SN~1998bu with the facility ``Tek2048 \#3''
CCD on the 0.9m telescope at CTIO on 16 May 1998 and finished on 14
July when the supernova reached the west limit of the telescope at
twilight. We obtained data on 30 nights, often through cirrus due to
the poor weather associated with the 1998 El Ni\~no weather pattern.
We also obtained a small amount of data with other telescopes.  We
observed one night with the CTIO Curtis Schmidt telescope with the
``Tek2048 \#5''CCD and the facility filters. We observed two nights
with the Las Campanas Observatory (LCO) Swope 1.0m telescope ``2x3K
SITe\#3'' CCD and facility filters.

The CCD at the 0.9m telescope is a thinned, anti-reflection coated
front-side illuminated CCD with 0.024mm (0\farcs40) pixels. The CCD
has very low read noise at 4 e- through all four amplifiers and also
has very high full-well depths of about 150,000 e-. We observed the
supernova with the facility ``Tek \#2'' $UBV(RI)_{KC}$ filters. The
filters have the following prescriptions: $U$, UG1/1mm + WG295/1mm +
CuSO$_4$ (5mm cell with saturated liquid) ; $B$, GG385/1mm + BG1/2mm +
BG39/2mm; $V$, GG495/2mm + BG38/2mm + BG39/1mm; $R_{KC}$, OG570/3mm +
KG3/2mm ; $I_{KC}$, interference filter with $\lambda_c=8050{\rm
\AA}$, FWHM=1500\AA.  The filters glasses were chosen from the Schott
Glass Technologies catalog and the listed dimensions are the
thickness.  The CTIO $V(RI)_{KC}$ sensitivity functions are a good
match to the those published by \cite{Bes90}.\footnote{Note that
\cite{Bes90} defines the sensitivity function as the product of the
quantum efficiency of the detector+telescope, the filter transmission
curve, the atmospheric extinction, and a linearly increasing function
of wavelength.}  The $B$ filter is about 150\AA\ blue of the Bessell
$B$ filter. In the following text we will refer to the filters
$(RI)_{KC}$ as simply $RI$.

The $U$ sensitivity function is quite different between the CTIO $U$
and the ``$UX$'' curve listed by Bessell. The CTIO curve has the same
red cutoff, but the blue cutoff is about 200\AA~redward of the Bessell
$UX$ cutoff due to the rapid falloff of the CTIO CCD quantum
efficiency below 3800\AA. Because the CTIO $U$ sensitivity is very low
blueward of the Balmer jump, the transformation from the natural to
the standard system will be sensitive to gravity and metallicity
differences at a fixed $(U-B)$ between photometric standards. This
will appear as extra random scatter in the $U$ transformations. For a
non-stellar spectrum of a supernova, a simple transformation from the
natural to the standard system cannot be expected to reproduce
accurate colors of the SN on the Johnson $U$ system. A correction,
akin to the ``K-correction'' will need to be applied. We will discuss
this effect on $U$ photometry in a future paper.

\subsection{Observations and Reductions to Photometry of the Local Standards}

The data were reduced through flat-fielding using standard techniques.
For the CTIO data, we removed the ``shutter error,'' which is the
increment in time across the CCD relative to the requested exposure
time due the finite open and close time of the iris shutter In this
case, the shutter error is about +0.065 seconds at the center and
+0.020 seconds at the corners.  A two-dimensional ``shutter image''
was created by comparing long and short exposures of the dome. This
image was used to correct the shutter error to better than 0.005
seconds everywhere on the image.  The long and short images were also
used to identify bad CCD pixels and low-level traps. These were masked
out of the final image before the photometric measurements.

All the CCD frames were processed through the programs DAOPHOT and
ALLSTAR of \cite{Ste87} and \cite{Ste94}. These programs output
point-spread-function (psf) photometry of the stellar images. We also
corrected the aperture photometry to a digital aperture diameter of
14\arcsec\ using the program DAOGROW (\cite{Ste90}) to correspond to
the standard photometer aperture used by \cite{Lan72} and
\cite{Lan92}.

On 10 nights we have enough photometric standards to calculate the
color terms of the transformation from the natural to the standard
$UBVRI$ system for the 0.9m telescope system. We have used the
technique suggested by \cite{Har_etal81}. They recommend that each
night be independently reduced and that the specific color terms be
averaged across all the runs. Each night is then re-reduced to the
final photometry with the averaged color terms.  We also use a
transformation equation of the form $m=f(M,I,X,T)$ suggested by
\cite{Har_etal81} where $m$ is the observed magnitude, $M$ is the
tabulated magnitude, $I$ is the tabulated color index, $X$ is the
airmass, and T is the time during the night. The inclusion of the
variable T allows for small and linear changes in the extinction
during the night. Winter-time observations at CTIO rarely require this
correction and we found the linear coefficient of this variable
consistent with zero for all 10 nights. We dropped this term from the
function. We also tried a cross term of the form $XI$ which is the
``second-order'' extinction term. The averages for the 10 nights in
all colors were consistent with zero and this term was also
dropped. Finally, with a linear form of the function $f$ we examined
the residuals for evidence of higher order terms. Only in $U-B$ was
there evidence of non-linear residuals. The inclusion of a quadratic
term in $U-B$ barely lowered the final dispersion, so we kept only the
linear term in $U-B$.

We list the averaged color terms in Table \ref{t1}. As an example,
Table \ref{t1} indicates that $v = a_0 + V - 0.018*(B-V) + a_1*X$.
The averaged extinctions in Table \ref{t1} are also listed for
reference purposes: however, we used only the nightly extinctions in
the reductions. The photometry used the standards of \cite{Lan72} and
\cite{Lan92}.  A typical night had 4-8 photometric fields with a total
of $\sim35$ standards. The color range was usually $(-0.3,2.0)$ in
$(B-V)$.

The photometry programs keep track of the total variance from the
photon statistics, errors in the psf, and the errors in the tabulated
magnitudes of the standard stars. The final residuals in the
transformation are compared to the expected errors. For $BVRI$ the
measured and estimated standard deviations are the same to within
0\fm01.  For $U$, one must add about 0\fm035 in quadrature to the
estimated error to force agreement with the observed dispersion in the
residuals. This extra dispersion is another indication that the
transformation of the $U$ photometry is uncertain due to the
non-standard transmission function of the $U$ filter/CCD system.

With the color terms fixed, we measured the local standards near NGC
3368 marked in Figure \ref{f1} for 11 photometric
nights.\footnote{There was one more photometric night used here than
was used for creating the color terms. On this extra night only a few
standards of limited color range were observed.} The averaged colors
for the 15 local standards are given in Table \ref{t2} and represent
aperture photometry with a 14\arcsec\ digital aperture. The table also
lists the number of independent nights of data in each color. Except
for the fainter stars, the mean errors are below 0\fm01. There are
more nights of data for the stars near NGC 3368 because roughly half
the data were taken with only the central 1024x1024 pixels read
out. The standards 14 and 15 were added later, and the photometry of
these stars was measured relative to the local stars rather from the
full nightly photometric solutions. The fainter stars in the table
were not used in the subsequent measurements of SN~1998bu.

In Table \ref{t1} we list the color terms derived for the Schmidt data
based on one photometric night and for the LCO data based on two
photometric nights. For the Schmidt data, only a single Landolt field
was observed at various airmasses, and the color terms should be used
with caution. The errors in the color terms for the Schmidt data
represent the residuals from a linear fit to the color index and X
whereas the color term errors for the 0.9m data represent the observed
night-to-night variation in the color terms. The color terms for the
LCO data are the averages for the two nights.

\subsection{Photometry of SN~1998bu}

With well determined color terms and local standards, we can now
transform the psf photometry to the standard system. The psf
magnitudes were fit over a {\it radius} equal to the gaussian FWHM of
the stellar images which varied from frame to frame with the
seeing. The sky value for the psf fit was determined in the star
subtracted image from the pixel values out to a radius of 8.8\arcsec\
excluding the central few pixels at the coordinates of the star. Each
frame of the supernova field was inspected for the quality of the psf
fitting and sky determination around the supernova. We typically used
linearly or quadratically varying psf's to achieve the highest quality
fits.  Each star in the psf was inspected to throw out non-stellar
objects from the psf calculation.

The supernova magnitudes can be affected by the galaxy background,
especially in the case where the supernova is in a spiral arm. One
cannot accurately know the galaxian background until the supernova
fades. When the supernova is bright, as is the case with our data,
this effect is not usually a problem. The surface brightness of the
spiral arm near the supernova is $UBVRI = (20.3,21.8,21.0,20.4,19.7)$
mag sq-arcsecond. At this magnitude level, the errors introduced by
assuming a uniform galaxian brightness at the position of the
supernova are small. For instance, if the supernova happens to be in a
region of twice the galaxian brightness inside an aperture of
diameter 3.2\arcsec, the faintest magnitudes in Table \ref{t3} will
decrease by (0.13,0.03,0.02,0.02,0.03) magnitudes.

The final $UBVRI$ photometry of SN~1998bu is given in Table \ref{t3}.
The light curves are plotted in Figure \ref{f2} and the color curves
in Figure \ref{f3}. The errors in Table \ref{t3} combine the errors in
the psf fits, the errors in the color terms, and mean error of the
photometric zero point in each frame. In many cases these errors are
less than 0\fm01. We note that while the mean error of the photometric
zero point is typically less than 0\fm01, the typical scatter of the
standards around the mean is more like 0\fm01 - 0\fm015. The zero
point is more accurate because we have 4-10 standards per field. We
regard this typical scatter of 0\fm01 as a more realistic minimum
error of a single observation of the supernova.

Finally, we caution the readers that {\it the $U$ photometry should be
considered uncertain due to the non-standard $U$ filter+CCD
transmission. Systematic errors of a few 0\fm1 could be present
especially in the late-time data.} Such a systematic error due to a
non-standard bandpass was seen in the photometry of SN~1987A
(\cite{Sun_etal88,Men89,Ham_etal90}) where differences of 0\fm3 were
seen in $I$.

\section{Discussion \& Conclusions\label{s2}}

From Figure \ref{f2} we see that the supernova was caught before
maximum.  Peak magnitudes in $UBVRI$ were measured directly via
low-order polynomial fits to the photometry near maximum.  These fits
are summarized in the first three columns of Table \ref{t4}.  As seen
in other well-observed type~Ia supernovae (SNe~Ia), maximum light in
$U$ was reached $\sim$1-2 days before $B$ maximum, which, in turn,
occurred $\sim$1-2 days before the $V$ and $R$ maxima.  Maximum light
in $I$ clearly took place {\em before} $U$ maximum; our observations
seem to have started just before $I$ maximum. 

To measure the initial decline rate of the $B$ light curve, we employ
the parameter \dmm, defined by \cite{Phi93} as the amount in
magnitudes that the $B$ light curve decays in the first 15 days after
maximum light.  For well-observed events such as SN~1998bu, \dmm\ can
be measured directly from a polynomial fit to the photometry.  We find
\dmm = 1.01 $\pm$ 0.05, close to the preliminary estimates of
\cite{Sch98}.

The red color ($B - V \sim$ 0.3) of SN~1998bu at maximum is indicative
of significant dust reddening, consistent with the strong \ion{Na}{1}
D and \ion{Ca}{2} interstellar absorption lines observed by
\cite{Mun_etal98} and \cite{Cen_etal98}.  From the dust maps of
\cite{Schl_etal98}, the Galactic reddening is estimated to be E($B-V$)
= 0.025 $\pm$ 0.003.  Hence, the majority of the reddening must have
been produced in the host galaxy of SN~1998bu, NGC~3368. Using the
methodology of Phillips et al. (1998), which is based on the B-V and
V-I color evolution of the SN, and assuming a Galactic reddening equal
to the Schlegel et al. value, we derive a host galaxy reddening of
E(B-V) = 0.34 +/- 0.03.  We note that this value is more than twice
the amount estimated by \cite{Mun_etal98} from the equivalent widths
of the \ion{Na}{1} D lines. The \cite{MunZwi97} empirical calibration
assumes a single-line component to the \ion{Na}{1} D equivalent
width. Multiple components have the effect of desaturating the line,
which means that reddening based on a given equivalent width for the
single-line component analysis should be an {\it upper limit}. This is
inconsistent with the reddening measured from the supernova colors. We
await the publication of the interstellar line data to explore this
matter further.

In the fourth column of Table \ref{t4}, the reddening-corrected peak
magnitudes of SN~1998bu are given assuming a standard Galactic
reddening law.  The distance to NGC~3368 (M96) has been measured by
\cite{Tan_etal95} from Cepheid variables discovered and observed with
the Hubble Space Telescope (HST).  These authors derived a {\it true}
distance modulus of (m-M)$_o$ = 30.32 $\pm$ 0.16 mag.  To this value,
an offset of +0\fm05 should be added to correct for the ``long''
versus ``short'' exposure zero point [see \cite{Sil_etal98} and
references therein].  Combining this distance with our
reddening-corrected peak magnitudes for SN~1998bu gives the absolute
magnitudes listed in the final column of Table \ref{t4}.

In Table \ref{t5}, which is an updated version of Table 2 from
\cite{Ham_etal96b}, we summarize the data for the four other
well-observed SNe~Ia with HST Cepheid-derived host galaxy distances,
as well as for three SNe whose host galaxies are members of a group or
cluster for which at least one HST Cepheid distance is available.  By
``well-observed'', we mean those SNe for which the photometry are
sufficiently precise to allow derivation of a reliable value of the
decline rate parameter \dmm.  For completeness, the data for SN~1998bu
are repeated in Table \ref{t5}, which has the following format:

Column(1): SN name.

Column(2): the name of the host galaxy.

Column(3): the true distance modulus of the host galaxy, or for the
group/cluster to which the host galaxy belongs. 

Column(4): references to the distance moduli given in column(3).

Columns (5), (6), and (7): the apparent $BVI$ peak magnitudes
of the SN as determined from the light curves.

Column (8): the observed value of the decline rate parameter 
\dmm.

Column (9): references to the SN photometry.

Column (10): the Galactic reddening E($B-V$) derived from the dust maps
of \cite{Schl_etal98}.

Column (11): the host galaxy reddening E($B-V$) as given by \cite{Phi_etal98}.

Columns (12), (13), and (14): the absolute $BVI$ peak magnitudes
of the SN as calculated from the apparent magnitudes, the distance modulus
of the host galaxy, and the Galactic and host galaxy reddenings.

There is now abundant evidence (see \cite{Ham_etal96a} and references
therein) that the absolute luminosities of SNe~Ia are a function of
the decline rate parameter \dmm.  Hence, to derive the
Hubble constant, \ho, from SN~1998bu and the other SNe listed in 
Table \ref{t5} we must use Hubble relations corrected for this effect.  
We assume the reddening-corrected fits given by \cite{Phi_etal98},
which lead to the following expressions for \ho\ in $BVI$:

\begin{displaymath}
{\rm log}\ho(B)=0.2\{M^B_{max}-0.720[\dmm-1.1]-1.010[\dmm-1.1]^2+28.653\},
\end{displaymath}

\begin{displaymath}
{\rm log}\ho(V)=0.2\{M^V_{max}-0.672[\dmm-1.1]-0.633[\dmm-1.1]^2+28.590\},
\end{displaymath}

\begin{displaymath}
{\rm log}\ho(I)=0.2\{M^I_{max}-0.853[\dmm-1.1]+28.219\}.
\end{displaymath}

\noindent
It should be noted that the observed decline rate of a SN~Ia is a weak
function of the dust extinction which affects the light curves
(\cite{Lei88}), and so the values of \dmm\ used in the above equations
should be corrected for this effect.  We use the following approximate
relation given by \cite{Phi_etal98}:

\begin{displaymath}
\dmm_{true} \simeq \dmm_{obs} - 0.1{\rm E}(B-V).
\end{displaymath}

Table \ref{t6} lists the resulting values of \ho\ for each of the SNe
with a Cepheid distance.  Note that the errors in this table are
internal -- i.e., they include the error contributions due to the
apparent magnitude measurements, reddening estimates, distance
modulus, absolute magnitude versus  decline rate relation, Hubble diagram
zero point, and the dispersion in the Hubble diagram only.  To these
must be added the external uncertainty in the zero point of the
Cepheid calibration, which we take to be $\pm$0\fm12 ($\pm$3.5 \kms\
Mpc$^{-1}$ in terms of \ho).

Table \ref{t6} shows that the five SNe with HST Cepheid distances give
results which are in excellent agreement, yielding a weighted average
value of \ho = 64.0 $\pm$ 2.2 \kms\ Mpc$^{-1}$, with an unweighted
dispersion of only 2.0 \kms\ Mpc$^{-1}$.  The three SNe with Cepheid
group/cluster show more scatter, yielding a somewhat higher weighted
average of 69.4 $\pm$ 3.3 \kms\ Mpc$^{-1}$.  A weighted average of all
eight SNe gives \ho = 65.6 $\pm$ 1.8 \kms\ Mpc$^{-1}$.  These values
are very similar to most other recent estimates of \ho\ based on
SNe~Ia (e.g.,
\cite{Ham_etal95,Rie_etal95,Ham_etal96a,Rie_etal96,Phi_etal98}).

Interestingly, the two most discrepant objects correspond to the Fornax 
cluster SNe, 1980N and 1992A, suggesting that the barred spiral galaxy 
NGC~1365 whose distance modulus we have taken to be that of the cluster 
may actually be foreground by $\sim$0\fm3.
This discrepancy is illustrated graphically in Figure \ref{f4} where we 
plot the absolute magnitudes given in Table \ref{t5} as
a function of the decline rate parameter \dmm.  
Also plotted as dotted crosses in Figure \ref{f4} are the 
reddening-corrected absolute magnitudes for 35 SNe~Ia in the Hubble flow 
(0.01 $\leq$ z $\leq$ 0.1) for which distances were calculated from the
observed radial velocities (corrected to the cosmic microwave background
reference frame) and an assumed Hubble constant of \ho = 65 \kms\ Mpc$^{-1}$.
The dashed lines show the fits derived by \cite{Phi_etal98} to the
latter data.  Note that both Fornax cluster SNe lie systematically lower 
than these fits by $\sim$0\fm3 in all three colors.  

According to \cite{Sil_etal98}, the HST Key Project on the Extragalactic
Distance Scale collaboration has recently observed two other spiral
galaxies considered to be members of the Fornax cluster, NGC~1425 and
NGC~1326A.  It will be very interesting to see if the Cepheid variables
discovered in these galaxies give true distance moduli closer to 31.6-31.7 as
suggested by the SNe data.  We also await with interest the results of
the HST observations announced by \cite{Sah_etal97} of the host galaxy of 
SN~1989B, NGC~3627, which should yield a direct measurement of the distance
to this SN.  Not only should these new distances provide more precise
estimates of the Hubble constant, but, as shown in Figure \ref{f4}, they 
will be extremely useful for confirming the shape of the absolute 
magnitude versus \dmm\ relation over a wide range of decline rates.

\acknowledgments

We would like to thank Dan Green and the Central Bureau for
Astronomical Telegrams which publishes the IAU Circulars for the fast
notification of this, and all, reported supernovae. We generally
receive notification within a day of discovery, which is critically
important for the organization of the followup observations.

{}

\newpage

\figcaption[nsuntzeff.fig1.ps]{SN 1998bu in M96 (NGC 3368). The local
standards listed in Table \ref{t2} are marked. This is a composite $V$
image taken on 17 May 1998 with the CTIO 0.9m telescope. The scale bar
corresponds to 1 arcminute. \label{f1}}

\figcaption[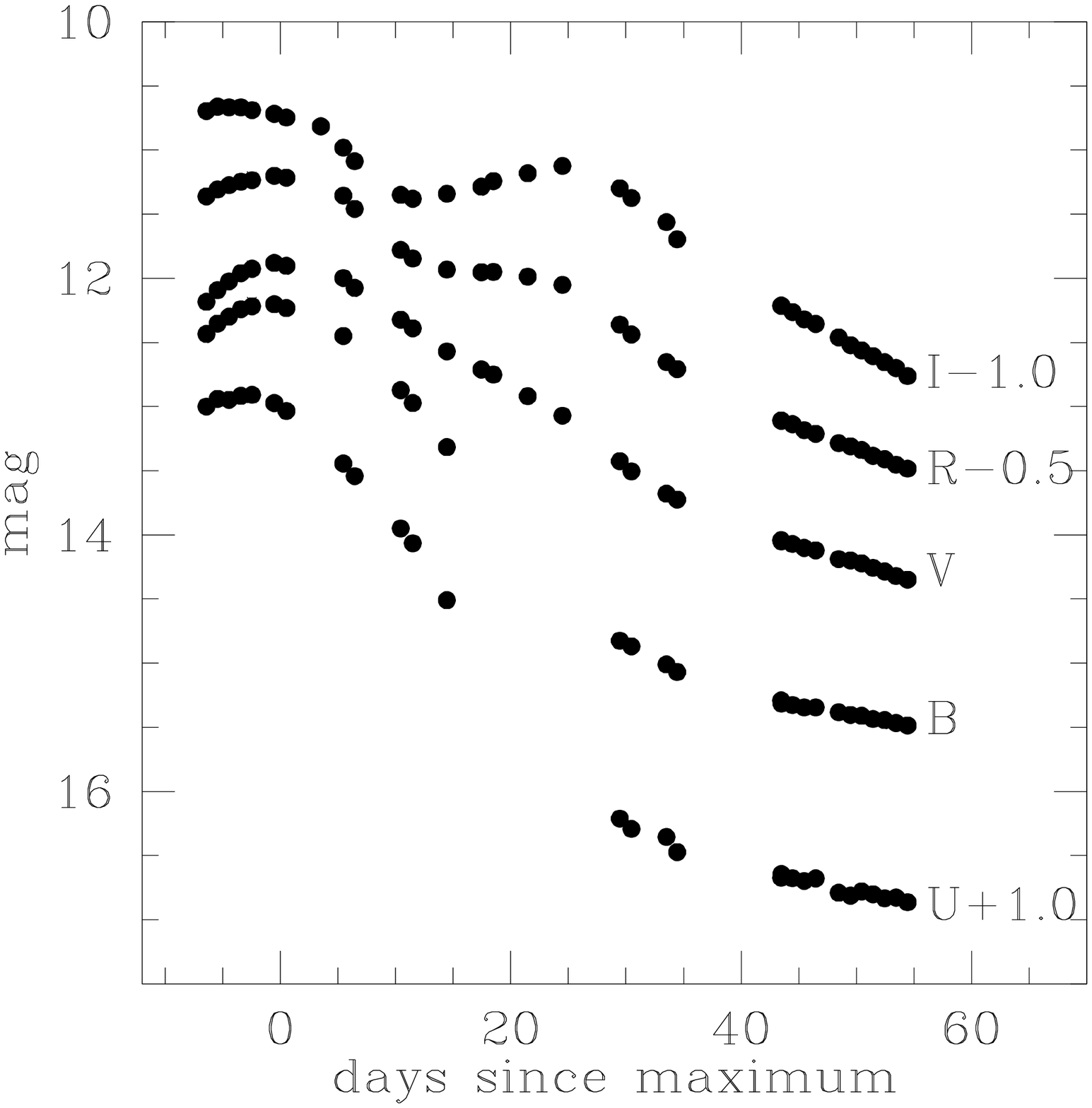]{The $UBVRI$ light curves for SN 1998bu
from Table \ref{t3}. The magnitude scales for $URI$ have been shifted
by the amount shown in the figure. The first two points are from data
taken at LCO. \label{f2}}

\figcaption[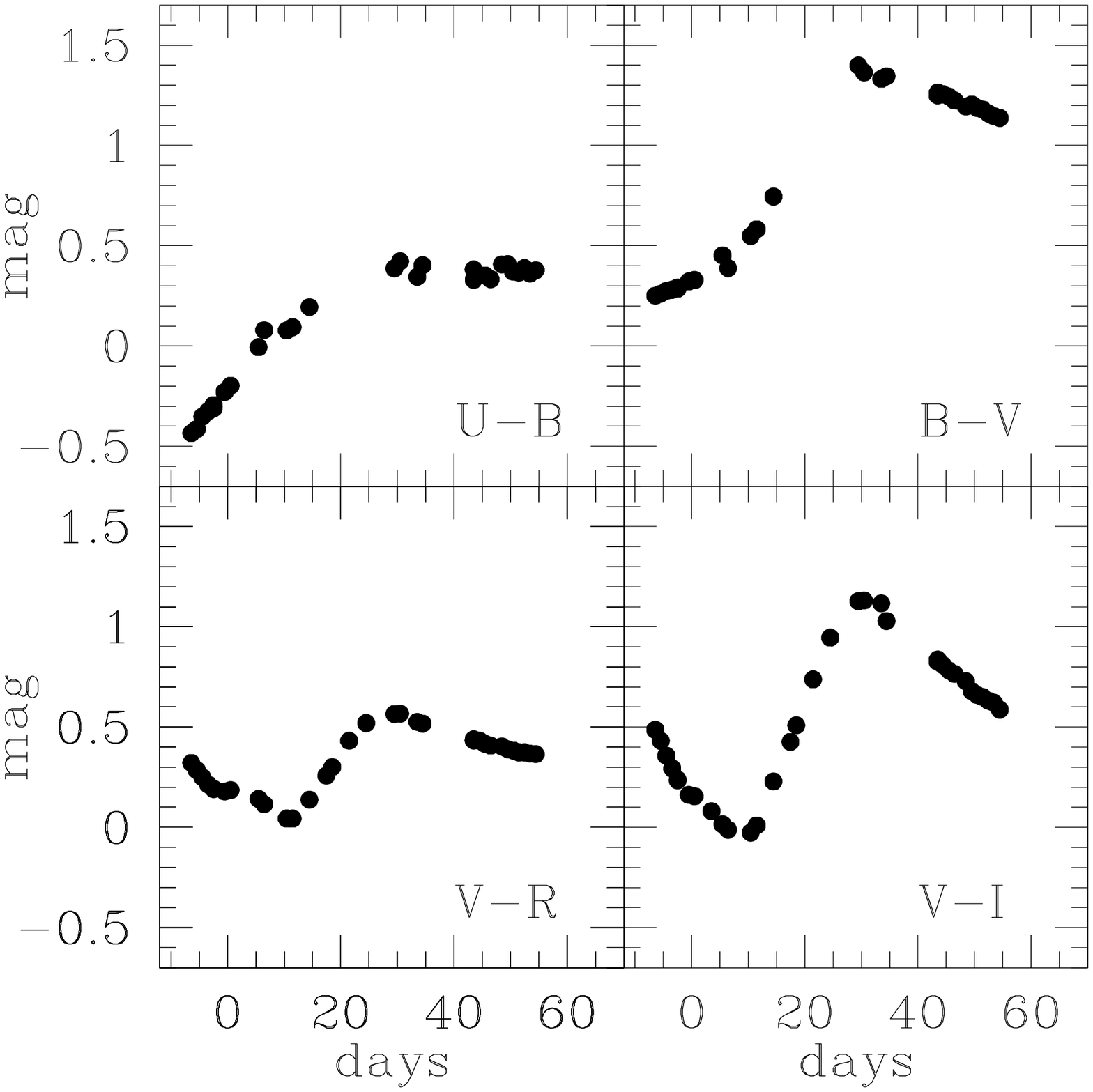]{The color curves for SN 1998bu from
Table \ref{t3}. The red colors at maximum light for a supernova with a
normal {\dmm} indicate a substantial reddening for this
object. \label{f3}}

\figcaption[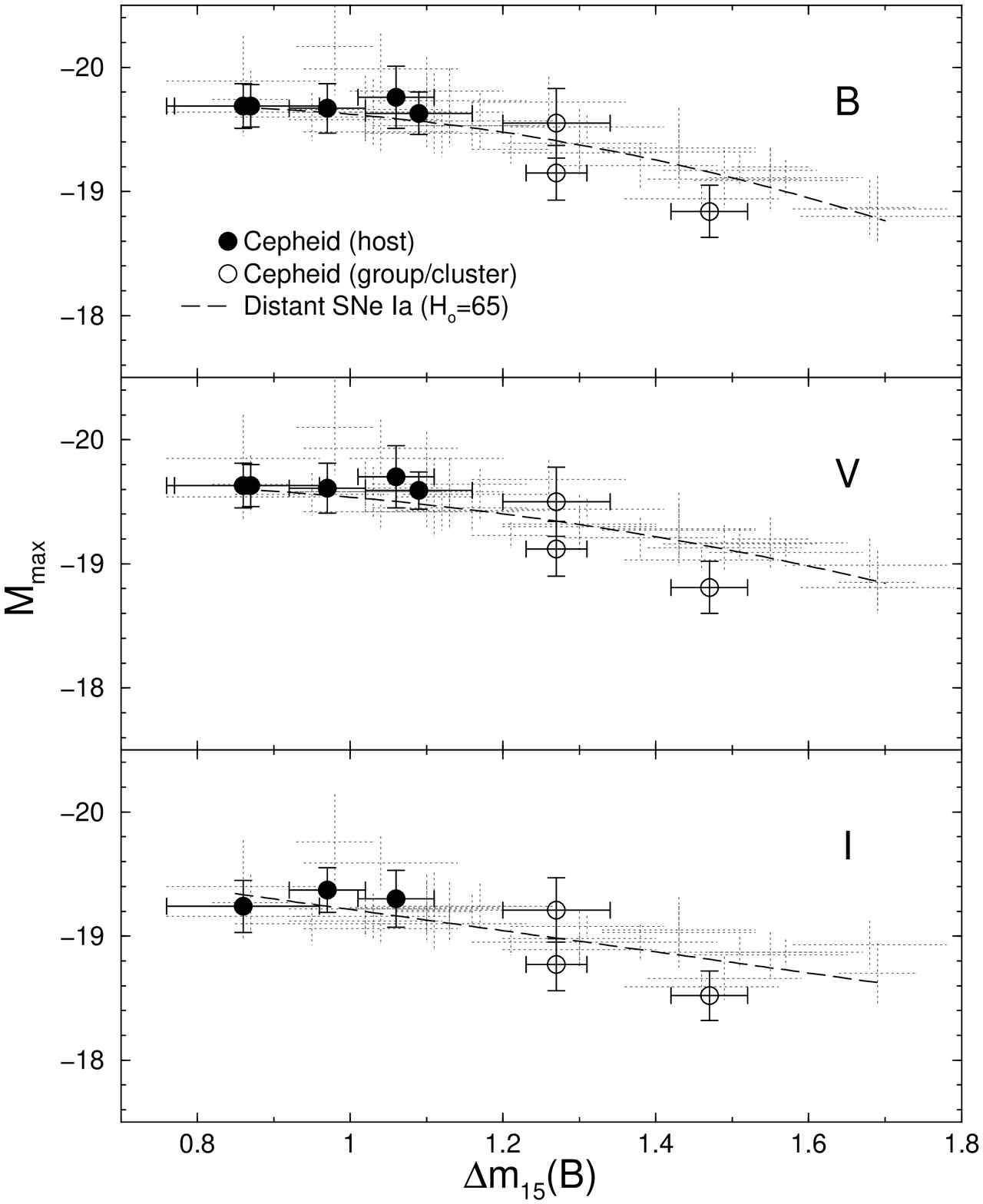]{The absolute magnitudes in $BVI$ for
Type~Ia supernovae as a function of the \cite{Phi93} parameter \dmm.
The filled circles correspond to the five SNe which occurred in host
galaxies with HST Cepheid distances. The open circles are SNe which
are members of groups or clusters for which at least one HST Cepheid
distance is available.  The dotted crosses are 35 SNe~Ia in the Hubble
flow whose distances were derived from their radial velocities and an
assumed Hubble constant of \ho = 65 \kms\ Mpc$^{-1}$, and the dashed
lines are fits to these data from \cite{Phi_etal98}.
\label{f4}}

\newpage

\begin{table}
\dummytable\label{t1}
\end{table}
{\sc TABLE} \ref{t1}. Averaged photometric transformation values

\begin{table}
\dummytable\label{t2}
\end{table}
{\sc TABLE} \ref{t2}. Photometry of the local standards near NGC 3368

\begin{table}
\dummytable\label{t3}
\end{table}
{\sc TABLE} \ref{t3}. Photometry of SN 1998bu in NGC 3368

\begin{table}
\dummytable\label{t4}
\end{table}
{\sc TABLE} \ref{t4}. Peak Photometric Magnitudes of SN~1998bu

\begin{table}
\dummytable\label{t5}
\end{table}
{\sc TABLE} \ref{t5}. Other SNe~Ia with HST Cepheid Distances

\begin{table}
\dummytable\label{t6}
\end{table}
{\sc TABLE} \ref{t6}. Values of H$_0$

\clearpage

  
\begin{figure}
\plotone{nsuntzeff.fig2.ps} 
{\center Suntzeff {\it et al.} Figure~\ref{f2}}
\end{figure}

\begin{figure}
\plotone{nsuntzeff.fig3.ps} 
{\center Suntzeff {\it et al.} Figure~\ref{f3}}
\end{figure}

\begin{figure}
\epsscale{0.6}
\plotone{nsuntzeff.fig4.ps} 
{\center Suntzeff {\it et al.} Figure~\ref{f4}}
\end{figure}

\clearpage


\begin{thebibliography}{}


\bibitem[Ayani et al. (1998)]{Aya_etal98} Ayani, K., Nakatani, H.,
 \& Yamaoka, H. 1998, \iaucirc\ 6905

\bibitem[Bessell (1990)]{Bes90} Bessell, M. S. 1990, \pasp, 102, 1181

\bibitem[Boschini 1998]{Bos98}  Boschini, L. 1998, \iaucirc\ 6902

\bibitem[Centurion et al. (1998)]{Cen_etal98} Centurion, M., 
 Bonifacio, P.,  Walton, N., \& King, D. 1998, \iaucirc\ 6918

\bibitem[Graham et al. (1997)]{Gra_etal97} Graham, J. A., Phelps, R. L.,
Freedman, W. L., Saha, A., Ferrarese, L., Stetson, P. B., Madore, B. F.,
Silbermann, N. A., Sakai, S., Kennicutt, R. C., Harding, P., Bresolin, F.,
Turner, A., Mould, J. R., Rawson, D. M., Ford, H. C., Hoessel, J. G.,
Han, M., Huchra, J. P., Macri, L. M., Huges, S. M., Illingworth, G. D.,
\& Kelson, D. D. 1997, \apj, 477, 535

\bibitem[Hamuy et al. 1990]{Ham_etal90} Hamuy, M., Suntzeff, N. B.,
Bravo, J., \& Phillips, M. M. 1990, \pasp, 102, 888

\bibitem[Hamuy et al. (1995)]{Ham_etal95} Hamuy, M., Phillips, M. M., Maza, J.,
Suntzeff, N. B., Schommer, R. A., \& Avil\'es, R. 1995, \aj, 109, 1

\bibitem[Hamuy {\it et al.} (1996a)]{Ham_etal96a} Hamuy, M., Phillips,
 M. M., Schommer, R. A., Suntzeff, N.B., Maza, J., \& Avil\'es, R.
 1996a, \aj, 112, 2391 

\bibitem[Hamuy {\it et al.} (1996b)]{Ham_etal96b} Hamuy, M.,
 Phillips, M. M., Suntzeff, N. B., Schommer, R. A., Maza, J., \&
 Avil\'es, R.  1996b, \aj, 112, 2398

\bibitem[Harris et al. (1981)]{Har_etal81} Harris, W. H., Fitzgerald,
M. P., \& Reed, B. C.  1981, \pasp, 93, 507

\bibitem[Landolt (1972)]{Lan72} Landolt, A. U. 1972, \aj, 78, 959

\bibitem[Landolt (1992)]{Lan92} Landolt, A. U. 1992, \aj, 104, 340

\bibitem[Leibundgut 1988]{Lei88} Leibundgut, B. 1988, Ph.D. Thesis, 
 Universitat Basel

\bibitem[Lira et al. (1998)]{Lir_etal98} Lira, P., Suntzeff, N. B., 
Phillips, M. M., Hamuy, M., Maza, J., Schommer, R. A., Smith, R. C., 
Wells, L., Avil\'es, R., Baldwin, J. A., Elias, J. H., Gonz\'alez, L.,
Layden, A., Navarrete, M., Ugarte, P., Walker, A. R., Williger, G. M.,
Baganoff, F. K., Crotts, A. P. S., Rich, R. M., Tyson, N. D., Dey, A.,
Guhathakurta, P., Hibbard, J., Kim, Y.-C., Rehner, D. M., Siciliano, E.,
Roth, J., Seitzer, P., \& Williams, T. B. 1998, \aj, 115, 234

\bibitem[Meikle et al. (1998)]{Mei_etal98}Meikle, P., Hernandez, M., \&
 Fassia, A. 1998, \iaucirc\ 6905

\bibitem[Menzies 1989]{Men89} Menzies, J.W., 1989, \mnras, 237, 21p

\bibitem[Munari et al. (1998)]{Mun_etal98} Munari, U., Barbon, R.,
 Tomasella, L., \& Rejkuba, M. 1998, \iaucirc\ 6902

\bibitem[Munari \& Zwitter (1997)]{MunZwi97} Munari, U., \&
 Zwitter. 1997, \aap, 318, 269

\bibitem[Perlmutter et al. (1997)]{Per_etal97}
 Perlmutter, S., Gabi, S., Goldhaber, G., Goobar, A., Groom, D. E.,
 Hook, I. M., Kim, A. G., Kim, M. Y., Lee, J. C., Pain, R.,
 Pennypacker, C. R., Small, I. A., Ellis, R. S., McMahon, R. G.,
 Boyle, B. J., Bunclark, P. S., Carter, D., Irwin, M. J., Glazebrook, K.,
 Newberg, H. J. M., Filippenko, A. V., Matheson, T., Dopita, M., 
 \& Couch, W. J. 1997, \apj, 483, 565.


\bibitem[Phillips (1993)]{Phi93} Phillips, M. M. 1993, \apjl, 413,
 L105

\bibitem[Phillips et al. (1998)]{Phi_etal98} Phillips, M. M., Lira, P.,
Suntzeff, N. B., Schommer, R. A., Hamuy, M., \& Maza, J. 1998, \aj,
submitted

\bibitem[Riess et al. (1995)]{Rie_etal95} Riess, A., Press, W. H., \&
Kirshner, R. P. 1995, \apj, 438, L17

\bibitem[Riess et al. (1996)]{Rie_etal96} Riess, A., Press, W. H., \&
Kirshner, R. P. 1996, \apj, 473, 88

\bibitem[Saha et al. (1994)]{Sah_etal94} Saha, A., Labhardt, L., 
Schwengeler, H., Macchetto, F. D., Panagia, N., Sandage, A., \& Tammann, G. A.
1994, \apj, 425, 14

\bibitem[Saha et al. (1995)]{Sah_etal95} Saha, A., Sandage, A., Labhardt, L.,
Schwengeler, H., Tammann, G. A., Panagia, N., \& Macchetto, F. D. 1995, 
\apj, 438, 8

\bibitem[Saha et al. (1996)]{Sah_etal96} Saha, A., Sandage, A., Labhardt, L.,
Tammann, G. A., Macchetto, F. D., \& Panagia, N. 1996, \apj, 466, 55

\bibitem[Saha et al. (1997)]{Sah_etal97} Saha, A., Sandage, A., Labhardt, L.,
Tammann, G. A., Macchetto, F. D., \& Panagia, N. 1997, \apj, 486, 1

\bibitem[Sandage (1961)]{San61} Sandage, A. R. 1961, \apj, 133, 355.

\bibitem[Schaefer (1998)]{Sch98} Schaefer, B. E., 1998, \iaucirc\ 6926.

\bibitem[Schlegel et al. (1998)]{Schl_etal98} Schlegel, D., Finkbeiner, D., 
\& Davis, M. 1998, \apj, in press

\bibitem[Schmidt et al. (1998)]{Schm_etal98} Schmidt, B. P., Suntzeff,
N. B., Phillips, M. M., Schommer, R. A., Clocchiatti,A., Kirshner,
R. P., Garnavich, P., Challis, P., Leibundgut, B., Spyromilio, J.,
Riess, A. G., Filippenko, A. V., Hamuy, M., Smith, R. C., Hogan, C.,
Stubbs, C., Diercks, A., Reiss, D., Gilliland, R., Tonry, J., Maza,
J., Dressler, A., Walsh, J., \& Ciardullo, R. 1998, \apj, 507, in
press.

\bibitem[Silbermann et al. (1998)]{Sil_etal98} Silbermann, N. A., et al. 1998,
preprint (astro-ph/9806017)

\bibitem[Skiff \& Faranda (1998)]{SkiFar98} Skiff, B.A., \& Faranda,
 C. Jr. 1998, \iaucirc\ 6905.

\bibitem[Stetson (1987)]{Ste87} Stetson, P. B. 1987, \pasp, 99, 191

\bibitem[Stetson (1990)]{Ste90} Stetson, P. B. 1990, \pasp, 102, 932

\bibitem[Stetson (1994)]{Ste94} Stetson, P. B. 1994, \pasp, 106, 250

\bibitem[Suntzeff et al. 1988]{Sun_etal88} Suntzeff, N.B., Hamuy, H.,
Martin, G., G\'omez, A., \& Gonz\'alez, R. 1988, \aj, 96, 1864

\bibitem[Tanvir et al. (1995)]{Tan_etal95} Tanvir, N.R., Shanks, T.,
Ferguson, H. C., \& Robinson, D. R. T. 1995, \nat, 377, 27

\bibitem[Villi 1998]{Vil98} Villi, M. 1998, \iaucirc\ 6899

\bibitem[Wells {\it et al.} 1994]{Wel_etal94} Wells, L. A., Phillips,
 M. M., Suntzeff, B., Heathcote, S. R., Hamuy, M., Navarrete, M.,
 Fernandez, M., Weller, W. G., Schommer, R. A., Kirshner, R. P.,
 Leibundgut, B., Willner, S. P., Peletier, S. P., Schlegel, E. M.,
 Wheeler, J. C., Harkness, R. P., Bell, D. J., Matthews, J. M.,
 Filippenko, A. V., Shields, J. C., Richmond, W., Jewitt, D., Luu, J.,
 Tran, H. D., Appleton, P. N., Robson, E. I., Tyson, J. A.,
 Guhathakurta, P., Eder, J. A., Bond, H. E., Potter, M., Veilleux, S.,
 Porter, A. C., Humphreys, R. M., Janes, K. A., Williams, T. B.,
 Costa, E., Ruiz, M. T., Lee, J. T., Lutz, J. H., Rich, R. M.,
 Winkler, P. F., \& Tyson, N. D.  1994, \aj, 108, 2233

\bibitem[Walker (1998)]{Wal98} Walker, A. R. 1998, ``Post Hipparcos
Cosmic Candles'', ed. F. Caputo \& A. Heck , (Kluwer Academic Publ.:
Dordrecht), in press



\end{thebibliography}
\end{document}